\documentclass[%
 reprint,
 amsmath,amssymb,
 aps,
]{revtex4-1}
\usepackage{physics}
\usepackage{amsmath}
\usepackage{graphicx}
\usepackage{dcolumn}
\usepackage{bm}

\newcommand{\be}{\begin{equation}}
\newcommand{\ee}{\end{equation}}
\newcommand{\bea}{\begin{eqnarray}}
\newcommand{\eea}{\end{eqnarray}}
\newcommand{\eqnarr}{\begin{eqnarray}}
\newcommand{\eqnend}{\end{eqnarray}}

\newcommand{\ti}[1]{\Tilde{#1}}
\DeclareMathSymbol{\umu}{\mathalpha}{operators}{0}
\usepackage[utf8]{inputenc}
\usepackage{physics}

\begin{document}

\preprint{APS/123-QED}

\title{Decay of Qubits under arbitrary space-time trajectories: \\ The Zeno \& Anti-Zeno Effects}

\author{Asad Hussain}
 \email{asadhussain625@gmail.com}
\author{Hamza Ahmed}%
 \email{hamza.ahmed95@gmail.com}
\affiliation{%
 School of Science \& Engineering - Lahore University Of Management Sciences (LUMS) \\
 Opposite Sector U, D.H.A, Lahore 54792, Pakistan
}%



\date{\today}

\begin{abstract}
Modeling an arbitrarily accelerating qubit as an open quantum system, we derive an exact solution for the pure-dephasing model ($\sigma_z$ coupling) under arbitrary qubit space-time trajectories, as well as general expressions for the survival probabilities of finite-length qubits interacting with a massless scalar field under $\sigma_x$ coupling (an Unruh-DeWitt detector) to second order. We follow the regularization scheme presented in \cite{Schlicht}, to allow a finite length quantum detector to couple to the massless scalar field. We compute the decay rate of the qubit in different coupling regimes, (pure dephasing, Unruh-DeWitt) and explore the Quantum Zeno (QZE) \& Quantum Anti Zeno Effect (QAZE) as the qubit decoheres under it's interaction with the environment. We compute results for four example trajectories: stationary, uniform acceleration, oscillation and circular orbits.
\end{abstract}

\pacs{Valid PACS appear here}

\maketitle
%
%
%
%
\section{\label{sec:level1}Introduction}
Often, in our efforts to control quantum systems, it is desired that once qubit states are prepared, they are prevented from decohering through their coupling with the environment. One way to control the decoherence is through repeated measurements of the system. The Quantum Zeno Effect (QZE) is an effect whereby quantum systems subjected to repeated measurements tend to remain in their initial states. The corresponding Quantum Anti-Zeno Effect (QAZE) refers to the speed up of decay of certain systems when measured repeatedly with some frequency \cite{Kofman2000}. For many such systems, the analysis of these effects has been explored at length in the literature \cite{18,16,14,2,20,Matsuzaki2010}, and for generic systems, a general framework was outlined in \cite{AdamFramework}.  As  the  prospect  of creating quantum computers becomes increasingly closer to realization, such explorations are essential in providing  a  better  understanding  of  how  to  control them.

On another front, through different motivations, W.G. Unruh found in  \cite{Unruh1976}, that a point-like detector, uniformly accelerating in a minkowski vacuum with acceleration parameter $a$, will respond exactly as if it was in a thermal bath of temperature $\frac{a}{2\pi}$, at least to first order in perturbation theory. Ever since the discovery of this theoretical effect many studies have been done in various directions to understand the response of such detectors \cite{OftenClick,Schlicht,Takagi,Dimitris}. Most of these explorations have been focused on uniformly accelerating, point like detectors, which became the source of many divergences when analyzing the amplitudes, and a large part of the literature is devoted to regularizing the Wightman functions on the scalar field along their path, and deriving analytically expressions from careful analysis of poles. \cite{Dimitris}


A more comprehensive way to explore the decay and decoherence of qubits is often done by using techniques from the field of Open Quantum systems, by using certain canonical models. For example, a two level system (TLS) interacting with bath of harmonic oscillators is a well studied system in the field of open quantum systems. The way these TLS's couple to the environment is neatly packaged into two pieces of information, the TLS coupling operator ($\sigma_x$ or $\sigma_z$) and the bath spectral density $J(\omega)$. 

In this paper, we will connect previous derivations considered for stationary systems to work for systems that move arbitrarily in spacetime, and look at it's impact on the Zeno \& Anti Zeno Effects. One thing of note will be the loss of the concept of a well defined spectral density $J(\omega)$, an important pillar in the framework considered in \cite{AdamFramework}.

Firstly, we will generalize the previous framework in \cite{AdamFramework} to work for the time dependent Hamiltonians that come from considering two level systems on accelerating world lines, and find the general formula for the decay of such systems to \textit{second order in perturbation theory}. 

Secondly, we generalize the well-known exact solution for the pure dephasing model to non-uniformly accelerating systems, and use \textit{Lorentz Invariant} regularized Wightman functions for \textit{smeared fields} as derived in \cite{Takagi,Schlicht}. These smeared fields allow us to not only consider detectors of finite size, but in fact, any small finite sized system that moves on spacetime.

Lastly, we compute the decay rate of different interesting trajectories: stationary, uniform acceleration, simple harmonic motion and circular motion. The last three will be explored in their respective relativistic regimes. Showing our results, we will identify the different Zeno \& Anti-Zeno regimes and provide some intuition about their relation to their path. 

\subsubsection*{Quantum Zeno \& Anti Zeno Effects}

For any small system with it's own Hamiltonian $H_F$, coupled to the environment with Hamiltonian $H_B$, through a coupling term $F\otimes B$, there will be a significant change in the evolution of the system compared to it's evolution without coupling. We consider initializing the system in the state $\ket{\psi_{0}}$, and doing $N$ projective measurements $P_{\psi_0} = \ket{\psi_0 (t)}\bra{\psi_0(t)}$ in time intervals of $t$, where $\psi_0(t)$ is the initial state evolved without environment interaction, or just the initial state in the interaction frame. \cite{Chaudhry2014,Matsuzaki2010} The probability for the state to be found in $\psi(t)$, \textit{with} environment interaction, after time $t$ is defined to be the survival probability of the system, $s(t)$. After $N$ projective measurements the probability of survival $S = [s(t)]^N$. This can be further written as $[s(t)]^N = e^{-\Gamma(t) N t}$. This allows us to consider the quantity 
\begin{equation}
    \Gamma(t) = -\frac{1}{t}\text{ln}[s(t)]
     \label{DecayR}
\end{equation} as the effective decay rate of the system. Regimes where the decay rate is an increasing function indicates a Zeno Regime, whereas the Anti-Zeno regime occurs when the decay rate is a decreasing function of time. \cite{AdamFramework}

\section{Setup}
We consider a small system of finite extent $\epsilon$, and set it to move on a space time trajectory given by the worldline $X^{\mu}(\tau)$. To get the Hamiltonian of a TLS coupled with a scalar field, we will first model a localized harmonic oscillator interacting with position coordinate $Q$, with the scalar field $\phi(x,t)$. 

The Lagrangian of the system is:

\begin{equation}
\begin{split}
S &= \int (\frac{1}{2}\dot{\phi}^2 - \frac{1}{2}(\nabla\phi)^2) d^4 x + c \int  m \phi[x^{\mu}(\tau)] Q d\tau  \\ &+ \int \frac{m}{2}(\frac{dQ}{d\tau})^2 - \frac{k}{2}Q^2  d\tau
\end{split}
\end{equation}

The system was solved in the appendix, and we find that the resultant Hamiltonian is:

\begin{align}
H = \Bigg( \frac{\hat{P}^2}{2m} + \frac{k \hat{Q}^2}{2} \Bigg)\dot{\tau}(t) - c m \hat{\phi}[x^{\mu}(t)]\hat{Q}\dot{\tau}(t) + \sum_{k}\omega_{k} \hat{b}^{\dagger}_k \hat{b}_k
\end{align}

\subsubsection*{A Generalization}

To generalize, when we want to write an accelerating system with it's own Hamiltonian $H_{F}$ interacting with a scalar field with some sort of operator product $F\otimes B$, where $F$ is a system operator and $B$ is an operator for a scalar field, \textit{that isn't dependent on the field momentum}, then  generally the Hamiltonian is:

\be
\label{GeneralHamiltonian}
\hat{H}(t) = \bigg( \hat{H}_{F} + F\otimes B(x(t)) \bigg)\dot{\tau}(t) + \sum_{k}\omega_{k} \hat{b}^{\dagger}_k \hat{b}_k
\ee

Notice that for systems that are not time dependent in the accelerating frame, the only explicit time dependence in the Hamiltonian comes from the $\dot{\tau}(t)$ term, there is also some dependence through the $\hat{\phi}[x(t)]$ term. 

\subsection*{Accelerating TLS coupled to scalar field}

We use the generalization \eqref{GeneralHamiltonian} and replace the harmonic oscillator with a TLS Hamiltonian. The coupling function $F$ will be $c \hat{\sigma_i}$ (where $i \in \{x,z\}$) and the TLS Hamiltonian will be:

\be \hat{H}_{TLS} = \omega_0 \hat{\sigma_z} + \Delta \hat{\sigma_x} \ee

The Boson Hamiltonian is:

\be
\hat{H}_{B} = \sum_{k}\omega_{k} \hat{b}^{\dagger}_k \hat{b}_k
\ee

Our coupling operators in the Schrodinger picture are:

\be
B(t) = \hat{\phi}[x(t)]  = \sum_{k} g^{*}_k (x(t))  \hat{b}_k + g_k (x(t)) \hat{b}_k^{\dagger}
\ee

We have written the $g_k(x(t))$ as $g_k(t)$ given that it is understood that $g_k(t)$ depends on implicitly depends on time through $\vec{x}(t)$

\be
F = c\hat{\sigma}_x \text{  or  } F = c\hat{\sigma}_z
\ee

And hence the total Hamiltonian for the system will be:

\begin{equation}
\begin{split}
\label{TLSHamiltonian}
\hat{H}(t) = \bigg(\omega \hat{\sigma_z} + \Delta \hat{\sigma_x} + c \hat{\sigma_i}  \otimes  \big(\sum_{k} g^{*}_k (x(t))  \hat{b}_k + g_k (x(t)) \hat{b}_k^{\dagger}\big) \bigg)  \\ \times\dot{\tau}(t) + \sum_{k}\omega_{k} \hat{b}^{\dagger}_k \hat{b}_k
\end{split}
\end{equation}

\subsection*{\label{Wight}Wightman Function Regularization}
As we will see in, \eqref{PerturbativeFormula} the survival probability of the qubit (which is a measure of it's deviation from it's evolution without the environment), needs $\langle \Tilde{B}(t_1)\Tilde{B}(t_2) \rangle$, where $t_1 > t_2$ .  This is the Wightman function of the bath environment variable. In this case specifically, we have $B(t) = \phi(x(t),t)$.  By moving into the interaction picture, we can see that:  

\be
\ti{B}(t) =  \sum_{k} \frac{e^{ik^{\mu}z_{\mu}}}{\sqrt{2\omega_k}}  \hat{b}_k + \frac{e^{-ik^{\mu}z_{\mu}}}{\sqrt{2\omega_k}}\hat{b}_k^{\dagger}
\ee

In this paper we will only consider a vacuum background, where $\rho_B = \ket{0_B}\bra{0_B}$ to focus more on the decoherence due to the Unruh effect  as opposed to the one that comes only through temperature. It was discussed in \cite{UnruhQuantumComputer} that temperatures determine the time scale of decay of any system through the thermal timescale $t \approx \frac{\hbar}{k_B T}$.  We will not go into the the thermal regime for the sake of clarity.

If we calculate the Wightman Function in the case of a massless scalar field, we get, before regularization:


\begin{equation*}
\bra{0_B}\phi(x)\phi(x')\ket{0_B} = \frac{1}{(2\pi)^3}\int\frac{d^3k}{2\omega_k}\;e^{ik^{\mu}(x_{\mu}-{x'}_{\mu})}\quad,
\end{equation*}

Of course, this is a completely oscillatory integral and one needs a regularization procedure. This has been attempted extensively in the the literature for varying cases \cite{Unruh1976,OftenClick,Schlicht,Takagi}. The standard method employed usually in QFT is to choose the minkowski time variable $t$ and add a small complex term $t \rightarrow t - i\epsilon$. The physical significance of the $\epsilon$ , in the context of accelerating qubits was pointed out by \cite{Takagi}. There he considered the field functions $\phi(x,t)$ appearing in $\bra{0_B}\phi(x)\phi(x')\ket{0_B}$, and considered them as \textit{smeared fields}. The way one does that is by defining new smeared fields $\phi_{\epsilon}(x,t)$ by convolving them with a smearing function as:

\begin{equation}
\phi_{\epsilon}(\tau) = \int  f_{\epsilon} (\xi) \phi(x-\xi,t) d^3 \xi 
\end{equation}
One can then use these smeared field operators to define the Wightman function. These smearing functions physically represent the extent of the detector, since a detector of finite size would not couple to the field at the position of it's center, but would couple at with the field over it's whole extent with different strengths. To approximate a point-like detector, $f_{\epsilon}(\xi)$ must be mostly localized near the origin, and become zero fast as they differ from it. There are many such smearing functions to choose from, but the one that gives the most tractable solution is one of the form \cite{Takagi}:

\begin{eqnarray}
\label{eq:det}
f_\epsilon(\xi)=\frac{1}{\pi^2} \frac{\epsilon}{(\xi^2+\epsilon^2)^2}
\end{eqnarray}

Here, $\epsilon$ is approximately the "width" of this distribution. Taking the limit as $\epsilon \rightarrow 0$ gives us the dirac delta function, and hence the fields for a pointlike detector. This gives the same regularization procedure as $t \rightarrow t - i\epsilon$, but with more physical motivation.  The Wightman functions that result from this procedure are \cite{Schlicht}:

\begin{eqnarray*}
\label{NonLorentz}
W_{\epsilon}(\tau,\tau') = \bra{0}\phi(x(\tau))\phi(x(\tau'))\ket{0} \\ \\ =
\frac{-1/4\pi^2}{\left(t(\tau)
-t(\tau')-i\epsilon\right)^2 - \left(x(\tau)-x(\tau')\right)^2}
\end{eqnarray*}

It was shown in \cite{Schlicht,Takagi} that the resultant Wightman function is not a Lorentz invariant quantity, since it depends on the choice of filtration of minkowski space into time slices. To remedy this, a Lorentz invariant regularization procedure was given in \cite{Schlicht} which, in essence, replaced the change $t \rightarrow t - i\epsilon$ with the change $x^{\mu} \rightarrow x^{\mu} + i\epsilon u^{\mu}$ where $u^{\mu}$ is the 4-velocity of the observer. The result of this paper was the regularized Wightman function:

\small
\begin{eqnarray}
\label{eq:meinW}
W_{\epsilon} (\tau,\tau') =
\frac{1/4\pi^2}{\bigl( 
x(\tau)-x(\tau') -i\epsilon({\dot{x}}(\tau)+
{\dot{x}(\tau'))\bigr)^2}}
\end{eqnarray}

This is the regularized Wightman function we will be using in this paper.

\section{Analytical Results}
\subsection*{Second Order Perturbative Expression}
We start with the Hamiltonian in \eqref{TLSHamiltonian} and go to the interaction picture, (denoted by a tilde). We denote our transformations as:

\be
\ti{B}(t) = e^{-i\hat{H_B}t} \big(\sum_{k} g^{*}_k (x(t))  \hat{b}_k + g_k (x(t)) \hat{b}_k^{\dagger}\big) \bigg) e^{i\hat{H_B}t}
\label{Bt}
\ee

\be
\ti{F}(t) = e^{-i\hat{H_F}\tau(t)}  c \hat{\sigma_i} e^{i\hat{H_F}\tau(t)}
\ee

The formula for the density matrix as a function of time (to second order in perturbation theory) of the resultant state can be obtained from \cite{AdamFramework}:

\begin{align}
\begin{split}
\rho_F(T) = U_F(T)\Bigg[ \rho^F_0 + \int_0^T dt_1 \dot{\tau}(t_1) \int_0^{t_1}  dt_2 \dot{\tau}(t_2) \\ \times\big\{ \langle \Tilde{B}(t_1)\Tilde{B}(t_2) \rangle_{\rho^B_0} [\ti{F}(t_2)\rho^F_0,\ti{F}(t_1)] + h.c.\big\}\Bigg]U^{\dagger}_F(T)
\end{split}
\end{align}

and so can the formula for the survival probability:

\begin{equation}
\begin{split}
s(T) = 1 - 2 \text{Re} \Bigg[\int_0^T dt_1 \dot{\tau}(t_1) \int_0^{t_1}  dt_2 \dot{\tau}(t_2) \\ \times\langle \Tilde{B}(t_1)\Tilde{B}(t_2) \rangle_{\rho^B_0}\text{Tr}_F \Big\{ P_{\Psi^\perp}[\ti{F}(t_2)\rho^F_0,\ti{F}(t_1)] \Big\} \Bigg]
\end{split}
\label{PerturbativeFormula}
\end{equation}

$P_{\Psi^\perp}$ is the projector orthogonal to the initial system state, and we have taken a massless scalar field where:

\be
g_k = \frac{e^{-i\vec{k}\cdot\vec{x(t)}}}{\sqrt{2\omega_k}}
\ee

We may take the initial condition as:

\be
\rho_0 = \rho^F_0\otimes\rho^B_0  =|\uparrow\rangle\langle\uparrow |\otimes |0_B\rangle\langle 0_B |
\ee

for the case when the system-bath coupling comes is through $\sigma_x$, and is 

\be
\rho_0 = \rho^F_0\otimes\rho^B_0  =
\frac{1}{2}(\ket{\uparrow}+\ket{\downarrow})(\bra{\uparrow}+\bra{\downarrow})\otimes \ket{0_B}\bra{0_B}
\ee

for the case when the system-bath coupling comes is through $\sigma_z$. As mentioned in \ref{Wight} we take the scalar field to be in the vacuum state in all cases, since we want to just look at decoherence specifically due to the Unruh effect and not those due to thermal fluctuations.

\subsection*{An Exact Solution}

In this section we will extend a well-known derivation for the exact solution of the pure dephasing model. We will follow the derivation blueprint in \cite{TOOQS}, but make appropriate generalizations to accelerating qubits where needed.
The Hamiltonian we will consider is the one where we will set $\Delta=0$:
\begin{equation}
\begin{split}
\hat{H}(t)=\bigg(\omega_0 \hat{\sigma}_{z} + c \hat{\sigma}_{z} \otimes  \big(\sum_{k} g_{k}^{*}(x(t))\hat{b}_{k}+g_{k}  (x(t))\hat{b}_{k}^{\dagger}\big)\bigg) \\ \times \dot{\tau}(t) + \sum_{k}\omega_{k} \hat{b}_{k}^{\dagger} \hat{b}_{k}
\end{split}
\end{equation}
The total density matrix given by $\hat{\rho}(t)$. Here, since $[H,\hat{\sigma}_z]=0$, it follows that the diagonal components of the system density matrix, $\hat{\rho}_{S}(t)$ :
\begin{equation}
\begin{split}
\rho_{11}&=\text{Tr}_{S,B}\big[\ket{1}\bra{1}\hat{\rho}(t)\big]=\bra{1}\hat{\rho}_{S}(t)\ket{1} \\
\rho_{00}&=\text{Tr}_{S,B}\big[\ket{0}\bra{0}\hat{\rho}(t)\big]=\bra{0}\hat{\rho}_{S}(t)\ket{0}
\end{split}
\label{densitym}
\end{equation}
are constant in time. To find the off-diagonal elements and hence the total system density matrix, we'll first write our interaction part of the Hamiltonian, given by $H_{I}(t)$, which is given by:
\begin{equation}
\hat{H}_{I}(t)=c \hat{\sigma}_{z} \otimes  \bigg(\sum_{k} g_{k}^{*}(x(t))\hat{b}_{k}+g_{k}  (x(t))\hat{b}_{k}^{\dagger}\bigg)\times \dot{\tau}(t)
\end{equation}

In the interaction picture:
\begin{equation}
\tilde{H}_{I}(t)=e^{i\hat{H}_{0}t}\hat{H}_{I}(t)e^{-i\hat{H}_{0}t}
\end{equation}
This becomes:
\begin{equation}
\hat{H}_{I}(t)=c \hat{\sigma}_{z} \otimes  \bigg(\sum_{k} g_{k}^{*}(x(t))\hat{b}_{k}e^{-i\omega_{k}t}+g_{k}  (x(t))\hat{b}_{k}^{\dagger}e^{i\omega_{k}t}\bigg)\times \dot{\tau}(t)
\end{equation}
Now, the unitary time evolution in the interaction picture is given by:
\begin{equation}
\hat{U}(t)=\mathcal{T}\text{exp}\bigg[-i\int_{0}^{t} ds\tilde{H}_{I}(s)\bigg]
\end{equation}
where $\mathcal{T}$ denotes the time ordering symbol. For the calculation in \eqref{asad}, we will need the commutator of the interaction picture Hamiltonian $\tilde{H}_{I}(t)$, with itself at different times. It can be shown to be given by:
\begin{equation}
\big[\tilde{H}_{I}(t_{1}),\tilde{H}_{I}(t_{2})\big] = -i\sum_{k}\frac{\dot{\tau}(t_{1})\dot{\tau}(t_{2})}{\omega_{k}}c^{2} \text{sin}\big( k_{\mu}x^{\mu}(t_{1}) - k_{\mu}x^{\mu}(t_{2}) \big)
\label{scope1}
\end{equation}
where:
\begin{equation}
k_{\mu}x^{\mu}(t)=\omega_{k}t-k_{i}x^{i}
\end{equation}
As we can see, this is just a c-number function. We can write our time-evolution operator as:
\begin{equation}
\begin{split}
\hat{U}(t) =\text{exp}\bigg[-\frac{1}{2}\int_{0}^{t} ds\int_{0}^{t} ds'\big[\tilde{H}_{I}(s),\tilde{H}_{I}(s')\big]\theta(s-s')\bigg] \\ \times\text{exp}\bigg[-i\int_{0}^{t} ds \ \tilde{H}_{I}(s)\bigg]
\end{split}
\label{asad}
\end{equation}
From ~\eqref{scope1}, we can see that the the first exponential in the above equation is just a time-dependent phase factor, and hence the evolution of the system is only governed by the second exponential, and so, our effective time-evolution operator is:
\begin{equation}
\begin{split}
\hat{U}(t)_{eff}= \exp\Bigg[\sum_{k}-i\sigma_{z}c\int^{t}_{0} ds \bigg(g_{k}^{*}(x(s))\hat{b}_{k}e^{-i\omega_{k}s} \\ + g_{k}  (x(s))\hat{b}_{k}^{\dagger}e^{i\omega_{k}s}\bigg)\times \dot{\tau}(s)\Bigg]
\end{split}
\end{equation}
We can write this succinctly as:
\begin{equation}
\hat{U}(t)_{eff}= \exp\Bigg[\sum_{k} \frac{c}{2} \sigma_{z} \bigg(\alpha_{k}(t)\hat{b}_{k}^{\dagger}-\alpha^{*}_{k}(t)  \hat{b}_{k}\bigg)\Bigg]
\end{equation}
where:
\begin{equation}
\alpha_{k}=-2 i\int _{0} ^{t} ds \  g_{k} (s) e^{i\omega_{k}s} \dot{\tau}(s)
\end{equation}
Now that we have our time-evolution operator, we can find the system density matrix. From here on, I will refer to $\hat{U}(t)_{eff}$ simply as $\hat{U}(t)$.

Let us consider an initial state, $\hat{\rho}(0)$ given by:
\begin{equation}
\hat{\rho}(0)=\hat{\rho}_{S}(0)\otimes\hat{\rho}_{B}
\end{equation}
 We'll take $\rho_{B}$ as a thermal state:
\begin{equation}
\hat{\rho}_{B}=\frac{1}{Z_{B}}e^{-\beta \hat{H}_{B}}
\end{equation}
where $\beta =1/k_{B}T$ and $\hat{H}_{B}=\sum_{k}\omega_{k} \hat{b}_{k}^{\dagger} \hat{b}_{k}$. 
We now need to determine the off-diagonal elements (or coherences). These are given by:
\begin{equation}
\rho_{ij}(t)=\bra{i}\hat{\rho}_{S}(t)\ket{j}=\bra{i}\text{Tr}_{B}\big[\hat{U}(t)\hat{\rho}(0)\hat{U}(t)^{\dagger}\big]\ket{j}
\label{scope2}
\end{equation}
We can write these as:
\begin{equation}
\rho_{10}(t)=\rho_{01}(t)=\rho_{10}(0)e^{\chi(t)}
\end{equation}
Using our $\hat{U}(t)$ and ~\eqref{scope2}, we find that:
\begin{equation}
\chi(t)=\sum_{k}\text{ln}\bigg\langle \text{exp}\big[\alpha_{k}(t)\hat{b}_{k}^{\dagger}-\alpha^{*}_{k}(t)  \hat{b}_{k}\big]\bigg \rangle
\end{equation}
The angular brackets denote the expectation value with respect to the thermal state $\hat{\rho}_{B}$. This expectation value is the Wigner characteristic function of bath mode $k$. It can be evaluated by noting that it's a Gaussian function. We thus find:
\begin{equation}
\bigg\langle \text{exp}\big[\alpha_{k}(t)\hat{b}_{k}^{\dagger}-\alpha^{*}_{k}(t)  \hat{b}_{k}\big]\bigg \rangle=\text{exp}\bigg[-\frac{1}{2} c^{2}|\alpha_{k}|^{2}\bigg\langle\{\hat{b}_{k},\hat{b}_{k}^{\dagger}\}\bigg \rangle\bigg]
\end{equation}
where the curly brackets denote the anti commutator. 

Now we have:
\begin{equation}
\chi(t)=-\sum_{k}\frac{1}{2} c^{2}|\alpha_{k}|^{2}\bigg\langle\{\hat{b}_{k},\hat{b}_{k}^{\dagger}\}\bigg \rangle=-\sum_{k}\frac{1}{2} c^{2}|\alpha_{k}|^{2} \ \text{coth}\big(\omega_{k}/2k_{B}T\big)
\label{Gamma}
\end{equation}
where:
\begin{equation}
|\alpha_{k}|^{2}= 4 \int _{0} ^{t}\int _{0} ^{t} dsds' \  g_{k} (s) g^{*}_{k}(s') e^{i\omega_{k}(s-s')} \dot{\tau}(s)\dot{\tau}(s')
\end{equation}
This can be simplified to:
\begin{equation}
|\alpha_{k}|^{2}= 4 \int _{0} ^{T(t)}\int _{0} ^{T(t)} d\tau d\tau' \  g_{k} (\tau) g^{*}_{k}(\tau') e^{i\omega_{k}(\tau-\tau')}
\end{equation}
Where $T(t) = \tau(t)$. Plugging this in ~\eqref{Gamma}, we get:
\begin{equation}
\label{decayrate}
\begin{split}
\chi(t)=-\sum_{k} 2 c^{2}\int _{0} ^{T(t)}\int _{0} ^{T(t)} d\tau d\tau' \  g_{k} (\tau) g^{*}_{k}(\tau') \\ \times e^{i\omega_{k}(\tau-\tau')} \ \text{coth}\big(\omega_{k}/2k_{B}T\big)
\end{split}
\end{equation}
Since we are considering a vacuum background, we take $T=0$ and turn the summation over $k$ into an integral. We can then replace the whole $k$ integral by the Wightman Function given in ~\eqref{eq:meinW}. Hence, we get:
\begin{equation}
\chi(t)=-2 c^{2}\int _{0} ^{T(t)}\int _{0} ^{T(t)} d\tau d\tau' \ W_{\epsilon} (\tau,\tau') \
\end{equation}
Now that we have $\chi(t)$, we can calculate the density matrix of the system as a function of time through \eqref{densitym}, and hence, calculate the survival probability of the system with $\sigma_z$. It is then a straight forward us of equation \eqref{DecayR} to find the decay rate for this system for any path in space-time.

\begin{figure}[t]
\centering
\includegraphics[width=0.5\textwidth]{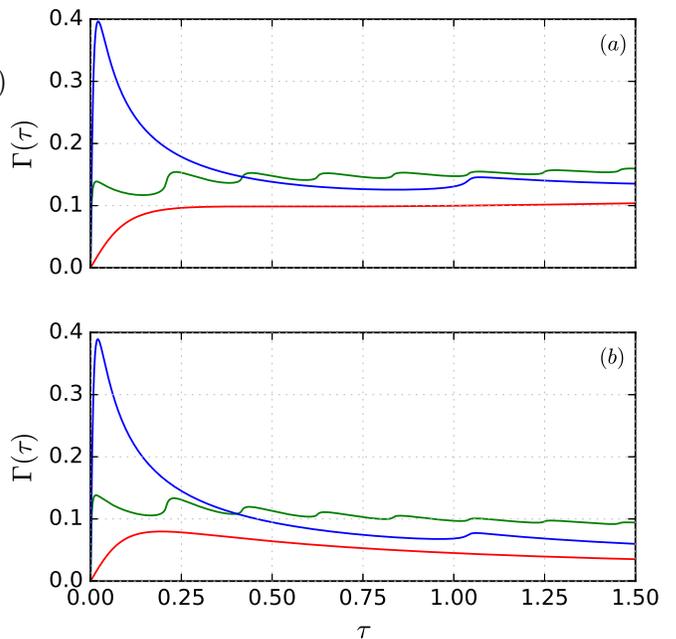}
\caption{\textbf{Oscillating Worldline}: $ (a) \ \Gamma(\tau)$ for $\sigma_x$,  and $ (b) \ \Gamma(\tau)$ for $\sigma_x$. \textbf{Red}: $\omega = 0\ eV$ (stationary). \textbf{Blue}: $\omega = 1.98\ eV$  \textbf{Green}: $\omega = 9.9\ eV$. All amplitudes are chosen such that the peak velocity during motion is $v = 0.99c$.}
\label{fig:SHM}
\end{figure}

\begin{figure}[t]
\centering
\includegraphics[width=0.5\textwidth]{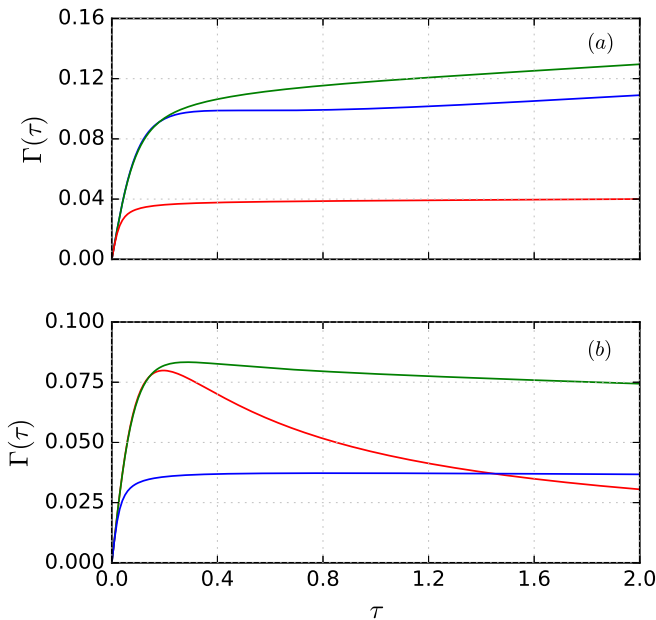}
\caption{\textbf{Uniformly Accelerating Worldline}: $(a):\ \Gamma(\tau)$ for $\sigma_x$ coupling. $(b):\ \Gamma(\tau)$ for $\sigma_z$ coupling. \textbf{(Red)} $a = 1\ eV$ \textbf{(Blue)} $a = 10\ eV$  \textbf{(Green)} $a = 100\ eV$.}
\label{fig:UA}
\end{figure}

\section{Results}
Our results in the previous section allow us to calculate effective decay rates for all possible worldlines. We plug in our path $X^{\mu}(\tau)$ into ~\eqref{eq:meinW}, from which we then calculate the decay rate. For illustrative purposes, we will consider four classes of worldlines that have been touched upon with varying degrees of importance in the literature \cite{CERNe,Experiment,OftenClick,Schlicht}, namely stationary, uniformly accelerating, oscillating and circularly orbiting worldlines. Also, throughout, we calculate the decay rate with respect to the proper time $(\tau)$ because for practical applications such as those in quantum computing, the computation itself takes place in the qubit rest frame \cite{ComputeRel}. In computing our results, we will work in natural units, and we set $\Delta = 0\ eV$ and $\omega_0 = 2\ eV$ for all cases.

\begin{figure}[t]
\includegraphics[width=0.5\textwidth]{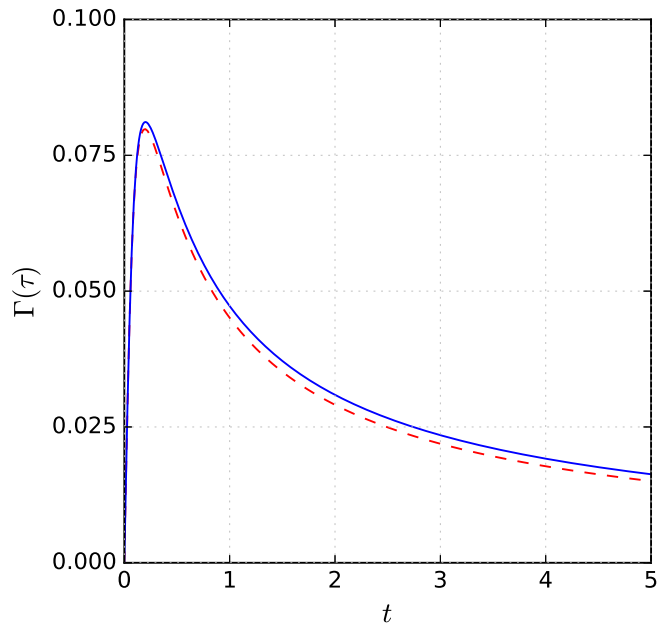}
\caption{\textit{Benchmarking}, $G=0.01$ and $\omega_{c}=10$.  \textbf{Blue}: Perturbative Solution. \textbf{Dashed Red}: Exact Solution.}
\label{fig:bm}
\end{figure}

First, we will consider a qubit on a stationary wordline, which will allow us to benchmark our results to that of previous calculations in relevant literature, and also allow us to make a link between the parameters in the spectral density $J(\omega)$ and the parameters of the qubit system. $J(\omega)$ usually has a few parameters depending on it's form, but to match it with our finite length qubit coupled with a scalar field, we will have an ohmic spectral density with an exponential cutoff, i.e. $J(\omega) = G \omega e^{-\frac{\omega}{\omega_c}}$ where $G$ is an overall coupling constant and $\omega_c$ is the cutoff frequency. The matching between our system parameters ($\epsilon$,$c$,$\omega_0$,$\Delta$) and the spectral density parameters are:
\begin{equation}
\begin{split}
\epsilon&=\frac{1}{2\omega_{c}} \\
c&=2\pi\sqrt{G}
\end{split}
\end{equation}
The remaining parameters effect the system's internal evolution. 
For the stationary case, we use the world line $X^{\mu}(\tau)=(\tau,0,0,0)$. Our result, given in Figure ~\ref{fig:bm}, is what we use to benchmark our exact solution (for $\sigma_{z}$ coupling) to our perturbative one (for $\sigma_{x}$ coupling); this is done for $G=0.01$ and $\omega_{c}=10$, as these parameters simultaneously allow us to match our results to the results in \cite{AdamFramework}. 

For the uniformly accelerating case, we have the worldline:

\begin{equation}
    X^{\mu} = \frac{1}{a}(\text{Cosh}(a\tau),\text{Sinh}(a\tau),0,0)
\end{equation}
where $a$ is the acceleration of the qubit.

The results for the computation for the worldline are in Figure \ref{fig:UA}, for both the $\sigma_x$ and $\sigma_z$ coupling regimes. An interesting thing to note from this result is that for the $\sigma_x$ coupling case, as $a$ increases, the decay rate increases and it seems that higher uniform acceleration inhibits Anti-Zeno regimes.

An oscillating qubit is given by the following worldline:
\begin{equation}
 X^{\mu} = (t,b \ \text{Sin}(\frac{v}{b} t),0,0)
\end{equation}
 where $b$ is the Amplitude of the oscillation and $v$ is the velocity which we have fixed at $0.99c$ to probe the relativistic regime. We see in Fig \ref{fig:SHM} that both the $\sigma_{z}$ coupling and the $\sigma_{x}$ coupling cases are similar. As for the trend, we notice that after the initial Zeno regime, the qubit stays in the Anti-Zeno regime until it finishes half its oscillation, after which there is a small Zeno regime, before it enters the Anti-Zeno regime again. For maximum coherence, as is desired in practical situations, one should match the frequency of measurement with the frequency of oscillation.

\begin{figure}[t]
\centering
\includegraphics[width=0.5\textwidth]{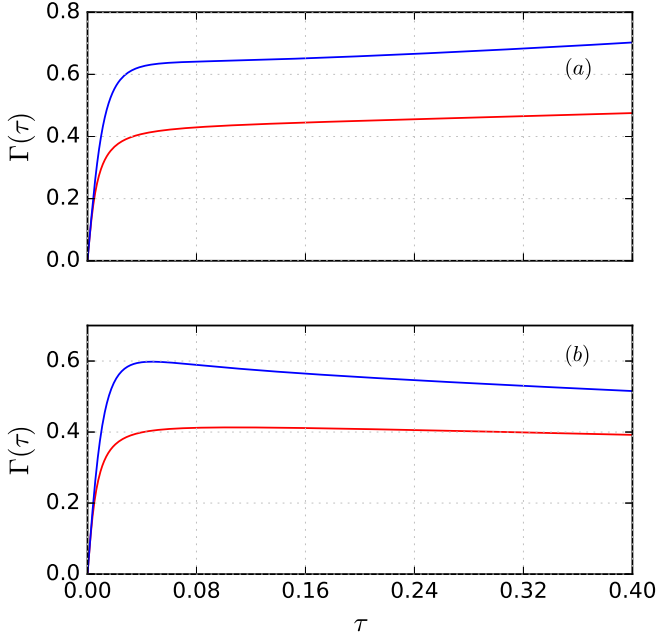}
\caption{\textbf{Circular Motion}: (a) $\Gamma(\tau)$ for $\sigma_x$ coupling, and (b) $\Gamma(\tau)$ for $\sigma_z$ coupling. \textbf{Blue}: $\omega = 1.98 eV$. \textbf{Red}: $\omega = 9.9eV$. All amplitudes are chosen such that the peak velocity during motion is $v = 0.99c$.}
\label{fig:CM}
\end{figure}

Our final case of Circular Motion is given by the worldline:
\begin{equation}
 X^{\mu} = (t,b \ \text{Sin}(\frac{v}{b} t),b \ \text{Cos}(\frac{v}{b} t),0)
\end{equation}
Here, $b$ is the radius of the circular orbit, and again, $v = 0.99c$. As illustrated in Figure \ref{fig:CM}, we see that the switching between Anti-Zeno and Zeno regimes (as in the oscillation case) disappears. In addition, the higher the radius of the circular motion, the more coherent our qubit remains.

\section{Discussion}
The physical realization of the setup we have proposed may be quite difficult. One clear reason comes down to pure dimensional analysis. It is well known that consequences of the Unruh Effect are difficult to discover, even though there have been some mechanism whereby the effect can be brought to the fore in circumstances achievable in particle accelerators \cite{CERNe}. Our analysis was done purely in natural units with the energy scale fixed to that of $eV$. Hence unruh effects clearly become significant at accelerations of $a = 1\ eV = 10^{23} ms^{-2}$, far beyond the scale of today's particle accelerators. In addition, though our chosen qubit length scale of $\approx 1\ eV^{-1} = 10^{-7} m$ seems close, our choice of oscillation frequency for our oscillating worldline, of the order of $\approx 1\ eV^{1} \approx 10^{16} Hz$, at a speed $0.99c$ is far too energetic to not be accompanied by a host of other complex reactions that would severely disrupt the experiment \cite{CERNe}. This problem comes down to the fundamental constants of our universe and not much can be done to overcome them. 

However, it is also important to make at least a heuristic argument to how the repeated measurements may be realized for a system that seems to be moving so rapidly. As long as the system's movement is localized enough to fit inside laboratory apparatus, one can propose a mechanism. Imagine coupling the system to another scalar field $\eta(x)$ (through the $\sigma_z$ operator) in a controlled way, and make the coupling much stronger than the coupling with $\phi(x)$. When a measurement needs to be made the coupling is immediately turned on, and the system completely decoheres on a time-scale much smaller than the time scale of the time evolution of the qubit. This constitutes a measurement of the system, since the qubit + $\eta$ field system becomes completely entangled like a pointer system and a measurement can be made on the radiation in $\eta$. However one can then only measure in the $\sigma_z$ basis, since pure dephasing results in final states of the form $\rho = a\ket{1}\bra{1} + b\ket{0}\bra{0} $. 

Let's assume we need to measure in a basis $\{\ket{b_i (t)}\}$, and we can relate this basis to the $\sigma_z$ eigenbasis by a unitary transformation $U(t)$. To remedy this, we first assume that we \textit{a priori} know $X^{\mu}(\tau)$. We can then extract the knowledge of the orientation of the spin vector of the particle, $\hat{\vec{S}}(\tau)$, and apply a time varying magnetic field to the particle such that $\hat{U}(t) = e^{-i \hat{\vec{S}}(t)\cdot \vec{B}(t)}$. We can apply this transformation very fast by a large magnetic field, and then immediately apply the measurement through the $\eta$ field. Subsequently, then transforming back to the original basis by $\hat{U}^{-1}(t)$ through the time-varying magnetic field. In such a way an experimental realization of this procedure may be possible, but more work is needed in such a direction.

\section{Acknowledgements}
We would like to thank Dr. Adam Zaman Chaudhry for his insight and guidance.

\appendix

\section{Hamiltonian for an Non-Uniformly Accelerated system}

\subsection{Unruh-DeWitt Detector System}
Before we consider the system above, we are going to derive the Hamiltonian for an accelerating harmonic oscillator. The reason for this is that the TLS has no Lagrangian as such, and no classical continuous configuration space. This doesn't allow us to canonically quantize properly. We will find out the Hamiltonian and then replace the harmonic oscillator system with the TLS. 

\subsection{Accelerating Harmonic Oscillator}
We consider two systems that are interacting, first, a harmonic oscillator which is moving through spacetime with some non-uniform motion $x^{\mu}(t)$ or $x^{\mu}(\tau)$ when parameterised by proper time, and second, a scalar field on flat minkowski space-time. It is key to note that the harmonic oscillator evolves normally along it's own world-line since the accelerating observer moving with the harmonic oscillator, just sees an ordinary harmonic oscillator.

The Lagrangian for this system can be given by:
\begin{align}
S = \int L_{\tt{field}}[\phi] + L_{\tt{interaction}}[\phi,Q]  + L_{\tt{Osc}}[Q] d\tau
\end{align}
\small

\begin{equation}
\begin{split}
S &= \int (\frac{1}{2}\dot{\phi}^2 - \frac{1}{2}(\nabla\phi)^2) d^4 x + c \int  m \phi[x^{\mu}(\tau)] Q d\tau  \\ &+ \int \frac{m}{2}(\frac{dQ}{d\tau})^2 - \frac{k}{2}Q^2  d\tau
\end{split}
\label{actionHO}
\end{equation}

Here the 'extension' of the Harmonic oscillator 'spring' is given by $Q$. It is worth noting that it does not extend in space, i.e. $Q$ has nothing to do with position in space, it is a completely localized system as far as our analysis is concerned. 

We can take the $t$ integral 'common' with the $\tau$ integral by:

\begin{equation}
\begin{split}
S = \int \int (\frac{1}{2}\dot{\phi}^2 - \frac{1}{2}(\nabla\phi)^2)) d^3 x dt + c \int  m \phi[x^{\mu}(\tau)] Q \frac{d\tau}{dt} dt \\ + \int \bigg( \frac{m}{2}(\frac{dQ}{d\tau})^2  - \frac{k}{2}Q^2   \bigg)\frac{d\tau}{dt} dt
\end{split}
\end{equation}

\begin{equation}
\begin{split}
S = \int \Big[ \int (\frac{1}{2}\dot{\phi}^2 - \frac{1}{2}(\nabla\phi)^2)) d^3 x + c m \phi[x^{\mu}(\tau(t))] Q \frac{d\tau}{dt} \\ + \bigg( \frac{m}{2}(\frac{dQ}{d\tau})^2  - \frac{k}{2}Q^2   \bigg)\frac{d\tau}{dt} \Big] dt
\end{split}
\end{equation}

\begin{equation}
\begin{split}
\implies L = \bigg( \frac{m}{2}\big(\frac{dt}{d\tau}\frac{dQ}{dt}\big)^2  - \frac{k}{2}Q^2   \bigg)\dot{\tau}(t) + c m \phi[x^{\mu}(t)] Q \dot{\tau}(t)  \\ +  \int (\frac{1}{2}\dot{\phi}^2 - \frac{1}{2}(\nabla\phi)^2)) d^3 x
\end{split}
\end{equation}

\begin{equation}
\begin{split}
\label{Lagrangian}
\implies L = \frac{m}{2\dot{\tau}(t)}\big(\dot{Q}\big)^2  - \frac{k}{2}Q^2\dot{\tau}(t) + c m \phi[x^{\mu}(t)] Q \dot{\tau}(t) \\ +  \int (\frac{1}{2}\dot{\phi}^2 - \frac{1}{2}(\nabla\phi)^2)) d^3 x
\end{split}
\end{equation}
Our configuration space consists of the field $\phi(x)$ and the harmonic oscillator's $Q$. The momentum conjugates are:

\begin{equation}
\pi(x) = \frac{\partial L}{\partial (\dot{\phi}(x))}
\end{equation}

\begin{equation}
\label{Pi}
\pi(x) = \dot{\phi}(x)
\end{equation}

and for the harmonic oscillator:

\be
P = \frac{\partial L}{\partial (\dot{Q})}
\ee

\be
\label{POsc}
P = \frac{m}{\dot{\tau}(t)} \dot{Q}
\ee

We can derive the Hamiltonian for the system by doing the legendre transformation:

\be
H = \int \dot{\phi}(x) \pi(x) d^3 x + P\dot{Q} - L
\ee

After substituting \ref{POsc}, \ref{Pi} and \ref{Lagrangian}, we get:

\begin{equation}
\begin{split}
H = \Bigg( \frac{P^2}{2m} + \frac{k Q^2}{2} \Bigg)\dot{\tau}(t) - c m \phi[x^{\mu}(t)] Q \dot{\tau}(t) \\ + \int \Bigg( \frac{{\pi(x)}^2}{2}  + \frac{(\nabla\phi)^2)}{2} \Bigg) d^3 x
\end{split}
\end{equation}

We can now promote them to operators. Now we can impose the commutation relations $[Q,P] = i\hbar$ and  $[\phi(x),\pi(x')] = i\hbar\delta(x-x')$. We now also expand the field into it's modes:

\begin{equation}
\begin{split}
H = \Bigg( \frac{\hat{P}^2}{2m} + \frac{k \hat{Q}^2}{2} \Bigg)\dot{\tau}(t) - c m \hat{\phi}[x^{\mu}(t)]\hat{Q}\dot{\tau}(t) \\ + \int \Bigg( \frac{{\hat{\pi}(x)}^2}{2} + \frac{(\nabla\hat{\phi})^2)}{2} \Bigg) d^3 x
\end{split}
\end{equation}

We can expand the field into it's modes and get:
\begin{align}
H = \Bigg( \frac{\hat{P}^2}{2m} + \frac{k \hat{Q}^2}{2} \Bigg)\dot{\tau}(t) - c m \hat{\phi}[x^{\mu}(t)]\hat{Q}\dot{\tau}(t) + \sum_{k}\omega_{k} \hat{b}^{\dagger}_k \hat{b}_k
\end{align}

\begin{align*}
\hat{H}(t) = \bigg( \hat{H}_{\tt{Osc}} + (-c\hat{Q})\otimes (m\hat{\phi}[x(t)]) \bigg)\dot{\tau}(t) + \sum_{k}\omega_{k} \hat{b}^{\dagger}_k \hat{b}_k
\end{align*}

We can expand $\hat{\phi}[x^{\mu}(t)]$ as:

\begin{equation} 
\label{ModeExpansion}
 \hat{\phi}[x^{\mu}(t)] = \sum_{k} \Big[ \hat{b}_k\frac{e^{i(k_i x^i (t))}}{\sqrt{2\omega_{k}}} + \hat{b}_k^{\dagger}\frac{e^{-i(k_i x^i (t))}}{\sqrt{2\omega_{k}}} \Big]
\end{equation}

We may define:

\be 
g_k = \frac{e^{-i(k_i x^i (t))}}{\sqrt{2\omega_{k}}}
\ee

and hence:

\begin{equation}
 \hat{\phi}[x^{\mu}(t)] = \sum_{k} g^{*}_k (x(t))  \hat{b}_k + g_k (x(t)) \hat{b}_k^{\dagger}
\end{equation}

\newpage
\bibliographystyle{plain}
\bibliography{bibliography.bib}

\end{document}